\documentclass[amsmath, amssymb,pra,preprint]{revtex4-1}
\usepackage{graphicx}
\usepackage{color}
\usepackage{mathrsfs}
\usepackage{lineno}

\begin{document}
\title{Photonic Crystal Optical Parametric Oscillator}
\author{Gabriel Marty$^{1,2}$ , Sylvain Combri\'{e}$^{1}$ , Fabrice Raineri$^{2,3}$ , Alfredo De Rossi$^{1,*}$}
\affiliation{$^{1}$ Thales Research and Technology, Campus Polytechnique, 1 avenue Augustin Fresnel, 91767 Palaiseau, France\\
$^{2}$Centre de Nanosciences et de Nanotetchnologies, CNRS, Universit\'{e} Paris Saclay, Palaiseau, France\\
$^{3}$Universit\'{e} de Paris, 5 Rue Thomas Mann, 75013 Paris, France\\
$^{*}$ Corresponding author: alfredo.derossi@thalesgroup.com}

\begin{abstract}
\textbf{Miniaturization of devices has been a primary objective in microelectronics and photonics for decades, aiming at denser integration, enhanced functionalities and drastic reduction of power consumption. Headway in nanophotonics is currently linked to the progress in concepts and technologies necessary for applications in information and communication\cite{sun2015}, brain inspired computing\cite{feldmann2019}, medicine and sensing\cite{estevez2012} and quantum information\cite{caspani2017}. Amongst all nanostructures, semiconductor photonic crystals (PhCs) \cite{akahane2003} occupy a prominent position as they enable the fabrication of quasi ultimate optical cavities\cite{asano2017}. Low threshold laser diodes\cite{nozaki2019,crosnier2017} or Raman lasers\cite{takahashi2013}, low power consuming optical memories\cite{nozaki2012}, efficient single photon sources \cite{lodahl2015} or single photon quantum gates\cite{Sun2018} are impressive examples of their capabilities.\\
We report the demonstration of a $\approx20\mu m$ long PhC semiconductor optical parametric oscillator (OPO) at telecom wavelength exploiting nearly diffraction limited optical modes. The pump power threshold is measured below 200$\mu$W. Parametric oscillation was reached through the drastic enhancement of Kerr optical Four Wave Mixing by thermally tuning the high Q modes of a nanocavity into a triply resonant configuration. Miniaturization of this paradigmatic source of coherent light paves the way for quantum optical circuits, dense integration of highly efficient nonlinear sources of squeezed light or entangled photons pairs.}
\end{abstract}

\maketitle

\begin{figure}[h!]
 \includegraphics[width=0.9\textwidth]{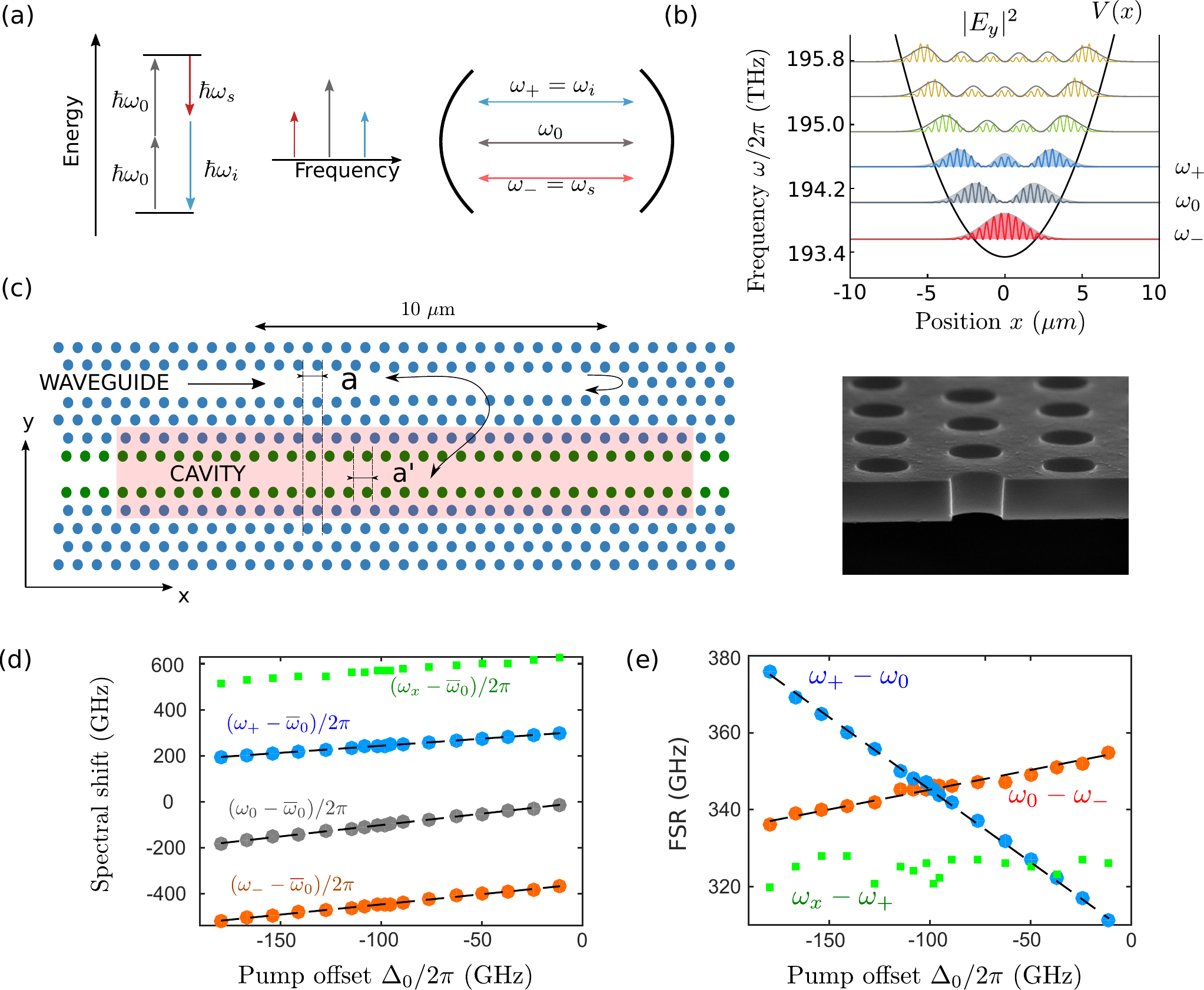}
\caption{\label{fig:cavity_tuning}  (a) Concept of the triply-resonant degenerate OPO where resonant modes respect the energy conservation $2\hbar\omega_0=\hbar\omega_+ + \hbar\omega_-$; (b) calculated Hermite-Gauss modes corresponding to an effective parabolic potential for photons (gray solid lines) and eigenmode frequencies and field amplitudes (color lines) of the bichromatic photonic crystal cavity; the filled curves denote a triplet of interacting modes; (c) Device layout with periods $a$ and $a^\prime$ and access waveguide and SEM image of the InGaP membrane;  (d) measured frequencies of the first 4 modes (relative to cold resonance $\overline\omega_0$) as a function of the pump offset $\Delta_0$ and linear fit (dashed); (e) corresponding eigenfrequency intervals (FSR)}
\end{figure}

Optical Parametric Oscillators (OPO) are sources of coherent light relying on the ultrafast nonlinear response of matter for the stimulated emission of photon pairs. OPOs can generate light within a spectral range only limited by the transparency of the nonlinear material. Consequently, they are broadly used in spectroscopy\cite{tittel2003}. As the generated photons are correlated, OPOs are also sources of nonclassical light for quantum optics\cite{Morin2012, morin2014} and quantum computing\cite{inagaki2016}.
OPOs are typically available as solid state optical devices, tending to be bulky and expensive. Their miniaturization is crucial for their deployment in photonic integrated circuits, e.g. for optical interconnects\cite{levy2010,razzari2010}. 
This challenging task can only be tackled by increasing power efficiency of the involved nonlinear effects in order to keep on-chip optical power as low as possible. Low power levels can be reached through the resonant enhancement of the interacting waves in high Q cavities. High-Q and low mode volumes are the two critical cavity parameters to decrease the power threshold for parametric oscillation\cite{matsko2005}.
The exceptional progress in the development of high-Q microresonators has enabled smaller and more practical OPOs\cite{kippenberg2004,grudinin2006}. In order to obtain even more compact devices, the use of nanocavities made of highly nonlinear material, such as semiconductors, seems to be the appropriate answer.
High-Q resonances with diffraction-limited mode volumes $\approx(\lambda/n)^3$, with $n$ the refractive index, are possible in a Photonic Crystal (PhC) cavity\cite{akahane2003,asano2017}. Although the possibility of a PhC OPO was considered theoretically more than a decade ago\cite{conti2004,ramirez2011}, the demonstration is extremely challenging and still missing.\\
\\
Our PhC OPO relies on harnessing triply resonant degenerate Four Wave Mixing (FWM) (see Fig. \ref{fig:cavity_tuning}(a)). For this Kerr optical nonlinear process, energy conservation dictates the relation between the pump and the generated photons ($2\omega_p=\omega_i + \omega_s$), meaning that the new frequencies $\omega_i$ and $\omega_s$ are generated symmetrically with respect to the pump one $\omega_p$. Thus, resonant enhancement is only possible if three cavity eigenfrequencies ($\omega_0$, $\omega_{+}$, $\omega_{-}$) are rigorously equispaced. This condition is in general not satisfied in PhC resonators. Here,  tight localization of light is related to the existence of forbidden bands in the optical spectrum\cite{john1987,yablonovitch1987}. In contrast to Fabry-Perot or ring resonators, PhC cavities can be designed to support a very small number of modes.  Triply resonant parametric interaction was observed in a system made of three coupled PhC nanocavities with a limited efficiency due to low Q factor\cite{azzini2013}. Similar experiments were performed in ultrahigh-Q coupled resonators waveguides\cite{matsuda2017} but did not lead to the expected drastic enhancement of the parametric interaction, likely because the eigenfrequencies were not equally spaced. In fact, the control of the frequency spacing is ultimately limited by structural disorder: even in state-of-the-art PhC technology, resonances deviate from their design value by about 40 GHz \cite{taguchi2011}. For this reason, in absence of a tuning mechanism, the eigenmode linewidth must be large enough to allow triply resonant FWM. This corresponds to $Q<5\times10^3$, clearly hindering the opportunity offered by PhC\cite{asano2017}.\\

In this work, we demonstrate a PhC OPO operated at ultra-low power. This result was achieved owing to three key features: first, we designed the cavity to have equispaced eigenfrequencies, then, we introduced a differential thermo-refractive tuning mechanism to compensate for the residual spectral misalignment caused by fabrication imperfections and, finally,we used In$_{0.5}$Ga$_{0.5}$P III- V semiconductor as the constitutive material of our cavity. The large electronic bandgap of this material enables the mitigation of Two Photon Absorption (TPA)\cite{colman2010} which otherwise clamps nonlinear conversion efficiency below parametric oscillations.
\\

The PhC cavity\cite{combrie2017} is designed to create an effective parabolic potential for the optical field, Fig. \ref{fig:cavity_tuning}(b), such that eigenfrequencies are equispaced and the eigenmodes correspond to Gauss-Hermite (GH) functions. The design (Fig. \ref{fig:cavity_tuning}(c)) consists of a suspended membrane with a regular hexagonal lattice of holes with period $a$, except for a line of missing holes, where light can propagate. There, we introduce a bichromatic lattice\cite{alpeggiani2015}, by modifying the period $a^{\prime}$ of the innermost row of holes. The localization of the field and the Free Spectral Range (FSR) between higher order modes are controlled by the commensurability parameter $a^{\prime}/a$. Choosing $a^{\prime}/a=0.98$ corresponds to a  FSR about 400 GHz and the field envelopes are very close to GH functions, Fig. \ref{fig:cavity_tuning}(b). The optical access to the modes is provided by a single-ended waveguide on the side terminated with a mode adapter to maximize coupling to an optical fiber\cite{tran2009}. \\


Our fabrication process ineluctably induces fluctuations on the targeted eigenfrequencies that result in a misalignment (50 GHz on average) far greater than the resonance linewidth (inferior to 1 GHz)\footnote{standard deviation over about 300 resonances in 72 resonators is 50 GHz, while linewidth of the resonances is as narrow as about 300 MHz, \cite{combrie2017}}. Thermal tuning can compensate for this deviation: a section of the cavity is heated to locally increase the refractive index of the material, hence the resonant wavelength of the modes\cite{Carmon2004}. When it is done on a smaller scale than the spatial extension of the targeted set of modes, each of them will spectrally  shift differently and relatively to their overlap with the temperature gradient. Local heating can be realized through the projection on the sample surface of a patterned incoherent pumping beam\cite{YuceLian2018}, this method being limited by the precision of the projection system. In our work, the tuning of three modes of a single cavity is automatically achieved by resonantly injecting light at mode "0", also used as the "pump" mode for the FWM process. A temperature gradient following the intensity  mode distribution is created inside the cavity through residual optical absorption. As can be seen in Fig. \ref{fig:cavity_tuning}(b), the GH modes exhibits different spatial profiles, inducing inhomogeneous spectral shifts. When the pump laser at $\omega_p$ is swept from blue to red  across the resonance, mode "0" redshifts from  $\overline{\omega_0}$ (cold) to  $\omega_0$ (hot). The spectral evolution of the pump, and the modes at $\omega_{+}$ and $\omega_{-}$ involved in the FWM process are represented in Fig.\ref{fig:cavity_tuning}(d). The different modes shift differently as expected. The misalignment of the hot cavity $2\Delta_\chi = 2\omega_{0}-\omega_--\omega_{+}$ is deduced from these measurements and is found to be $2\Delta_\chi = 2\overline\Delta_\chi + 0.48\Delta_0$ where  $\overline\Delta_\chi$ is the misalignment of the "cold" cavity and $\Delta_0 = \omega_p -\overline{\omega_0}$ is the pump offset.   This shows that an originally mismatched triplet can eventually be aligned by adjusting the pump frequency, which is clearly observed in Fig.\ref{fig:cavity_tuning}(e) where the two considered FSRs are equalized (crossing of blue and red lines). 

\begin{figure}[h!]
 \includegraphics[width=0.8\textwidth]{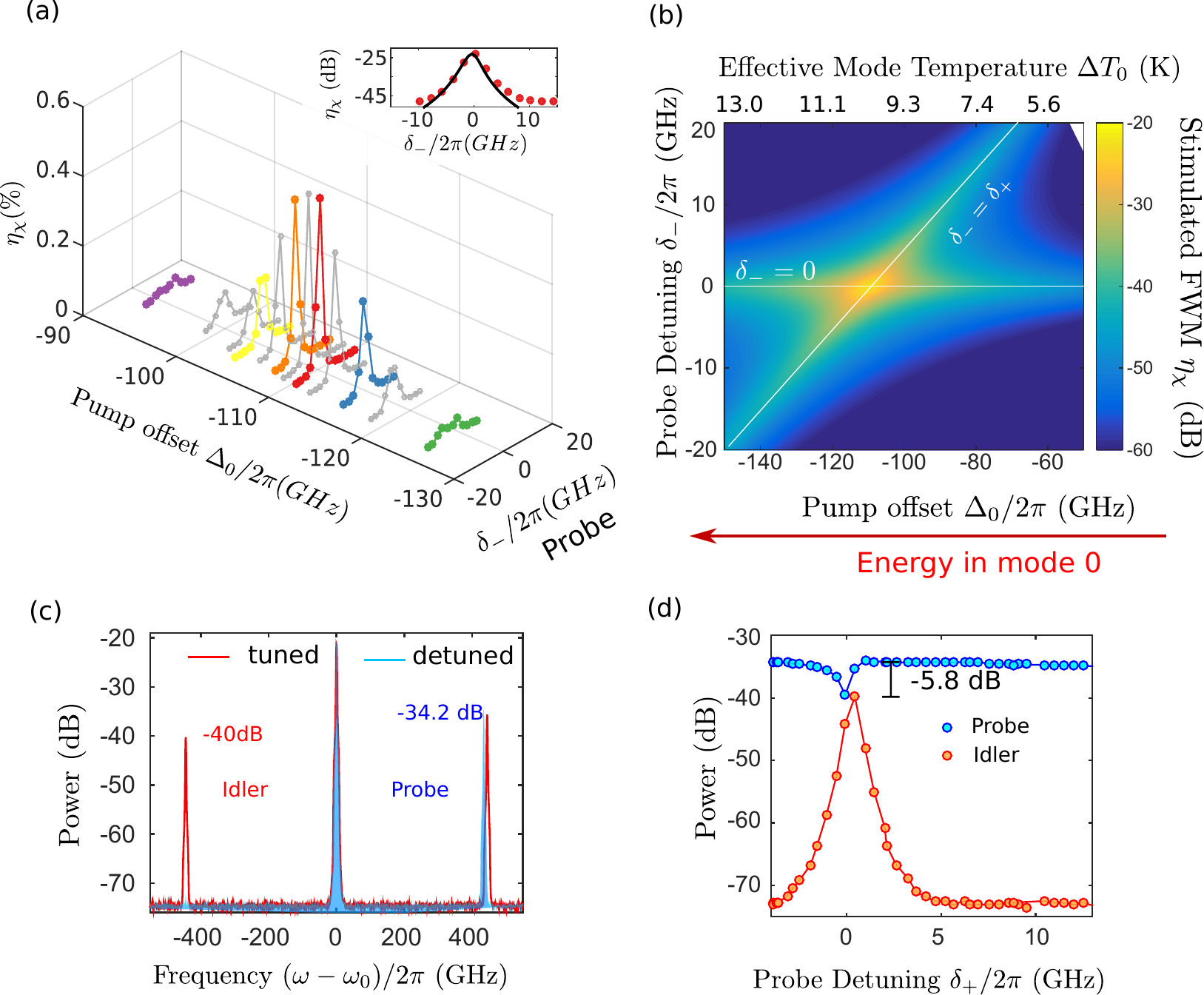}
\caption{\label{fig:FWM} (a) Measured stimulated FWM efficiency $\eta_\chi$ as a function of the pump offset $\Delta_0$ and probe detuning $\delta_-$;  (Inset) Comparison with the theory (black line) for the red  curve. (see Supplementary Fig.S2 for full comparison) (b) Calculated false color map of the efficiency $\eta_\chi$ of stimulated FWM  as a function of the probe detuning $\delta_-$ and  pump offset with corresponding effective mode temperature rise $\Delta T_0$ for mode 0. The white  lines represent the poles of eq. \ref{eq:FWM_lowgain_alt}.(c) Stimulated FWM in a resonator with larger $Q_{avg}$. Raw spectra centered on the pump $\omega_0$ as a function of the probe detuning $\delta_{-}$ spectra for tuned (red) and detuned (blue) probe; (d) reflected probe (blue) and idler (red) power vs. probe detuning.}
\end{figure}

We now investigate the impact of the triple resonance condition on the parametric gain. This dependence is deduced from stimulated FWM experiments, where an additional laser ("signal") at $\omega_s$ is swept across the resonance at $\omega_{-}$ when the pump frequency is fixed. An idler wave is generated at $\omega_i = 2\omega_p - \omega_s$.  As a matter of fact, the stimulated efficiency $\eta_\chi = P_i/P_s $, defined as the ratio between the output idler power ($P_i$) over the input signal power ($P_s$), is directly related to the parametric gain and the loss of the cavity. Parametric oscillation is reached when the gain is high enough to compensate for the loss, corresponding to the asymptotic limit of $\eta_\chi \rightarrow 1$.  The measured values of $\eta_{\chi}$ for a sample with an average Q\footnote{$Q_{avg}=(Q_0^2Q_- Q_{+})^{1/4}$}  of about $7 \times 10^4$ are displayed on Fig. \ref{fig:FWM}(a) as a function of the pump offset for an input pump power of 700 $\mu$W. $\eta_\chi$ peaks at 0.4 $\%$ (-25dB) when $\Delta_0/2\pi$ = -110GHz, in agreement with the triply resonant configuration predicted by our thermal tuning measurements shown Fig. \ref{fig:FWM}(b). 
To support these results, an analytical model (see Supplementary Information Section I) is derived and $\eta_\chi$ writes, in the limit of undepleted pump and low parametric gain, as: 

\begin{align}
\eta_\chi=\eta_\chi^{(max)}\mathcal{L}\left(\frac{\delta_{0}}{\Gamma_{0}}\right)^2\mathcal{L}\left(\frac{\delta_{-}}{\Gamma_{-}}\right)\mathcal{L}\left(\frac{\delta_{+}}{\Gamma_{+}}\right)
\label{eq:FWM_lowgain_alt}
\end{align}

where $\delta_0 = \omega_p-\omega_0$, $\delta_-= \omega_s - \omega_{-}$, $\delta_+= \omega_i - \omega_{+}$ are the frequency detunings of the pump,signal and idler with respect to the hot resonances.
 $\Gamma_0$ and $\Gamma_-$ , $\Gamma_+$ are the photon damping rates in the corresponding modes ($\Gamma = \omega/Q$). $\mathcal{L}(x)$ is the  Lorentzian function. $\eta_\chi^{(max)}$ is the maximum achievable efficiency obtained when the triplet is aligned and when the waves frequencies correspond to the resonant modes ($\delta_0 = \delta_- = \delta_+=0$). The calculated map of $\eta_\chi$ versus the pump offset and probe detuning $\delta_{-}$ reveals two local maxima merging into an absolute maximum of efficiency, corresponding to a perfectly aligned cavity (Fig. \ref{fig:FWM}(b)). Remarkably, the measurements are in quantitative agreement with our model (see Inset of Fig.\ref{fig:FWM}(a)) plugged with parameters either calculated or measured, and the intracavity power adjusted by less than 10$\%$ to account for experimental uncertainties (more details in the Supplementary Information Section II).


Thanks to our thermal tuning technique, higher FWM efficiencies are within reach by working with cavities exhibiting higher Qs even though the resonance linewidth become narrower. We performed stimulated FWM on a sample with an average $Q = 1.2\times10^5$.  It results in a drastic improvement of the efficiency that rises up to 26 $\%$ (– 5.8 dB) with an on chip pump power of 80 $\mu$W (Fig \ref{fig:FWM}(c)) and Fig \ref{fig:FWM}(d)). The scaling of efficiency with Q is detailed in the Supplementary Information Section I.(F). We also note that the dependence of the efficiency on the probe detuning, Fig. \ref{fig:FWM}(c),  is very sharp, because of the triple resonant enhancement of FWM. 

\begin{figure}
 \includegraphics[width=1.0\textwidth]{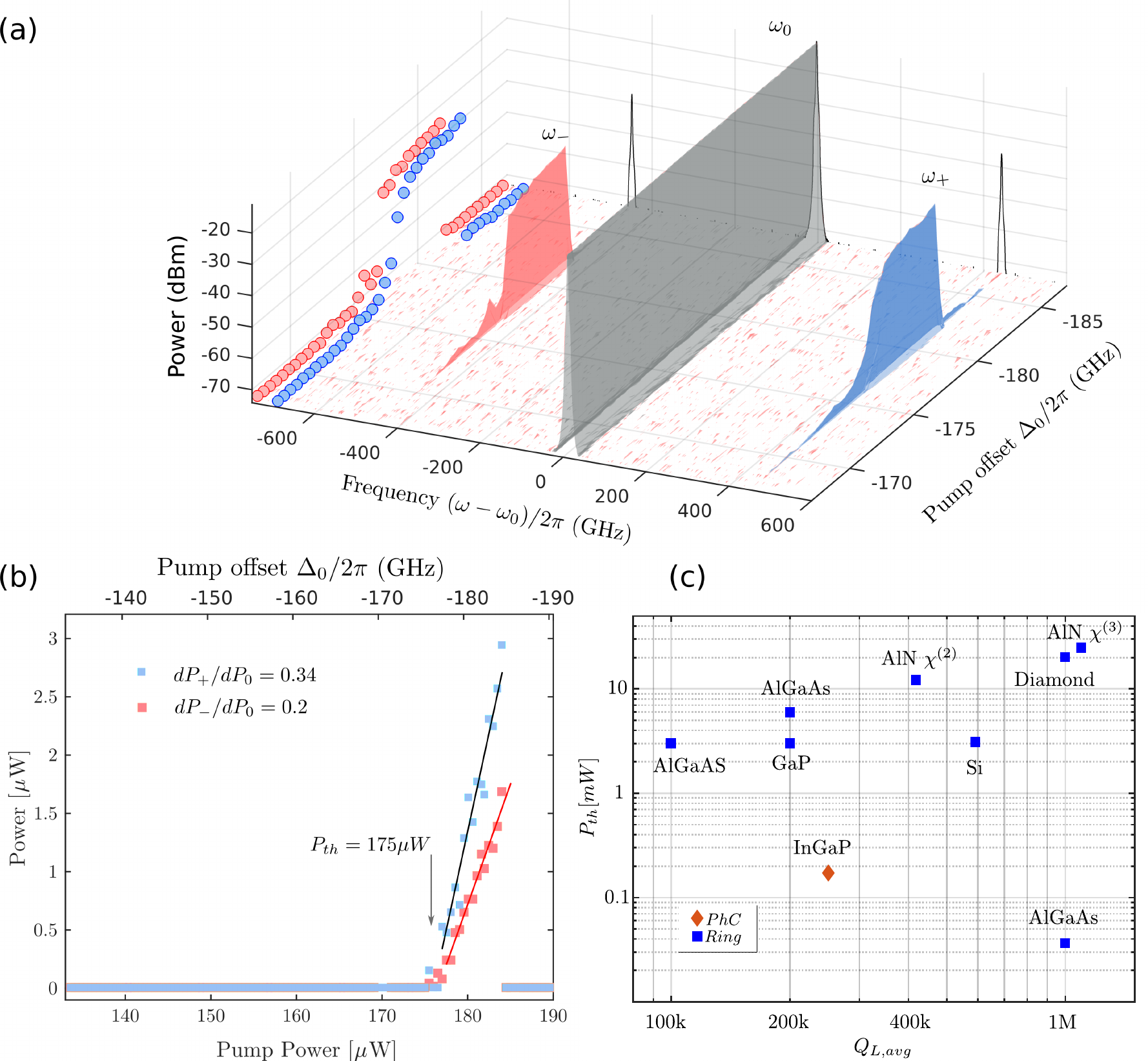}
\caption{\label{fig:OPO} 
(a) Parametric oscillation: raw optical spectrum (resolution 4 GHz, centered at the pump frequency) as the pump offset is changed. The threshold is overcome as the intra-cavity pump energy increases. Markers represent the raw power on the red and blue side, solid black line is the spectrum at maximum OPO emission $\Delta_0/2\pi=-183 GHz$; (b) On-chip power in the blue $\omega_+$ and red $\omega_-$ lines as a function of the pump offset and equivalent pump power in the cavity $P_{c,0}$; (c) OPO pump threshold as a function of the averaged Q factor $Q_{avg}$ in semiconductor integrated microring and racetrack resonators: AlGaAs\cite{pu2016,chang2019}, GaP\cite{wilson2019}, Diamond\cite{hausmann2014diamond},Silicon\cite{griffith2015}, AlN using $\chi^{(2)}$\cite{Bruch2019} and $\chi^{(3)}$\cite{liu2018integrated} nonlinearity.}
\end{figure}

%
%
%
In these samples, parametric oscillation is not observed, meaning that the maximum realizable parametric gain is not sufficient to compensate for cavity loss. This value is entirely determined by the cavity optical properties (Q, mode volume, nonlinear cross-section, material nonlinearity) and the pump power level required to align the three modes. Its value is therefore unique for each triplet of modes. 
Eventually, parametric oscillation is demonstrated for a resonator with $Q_{avg}\approx 2.5\times10^5$ . The sample is pumped with an on-chip power level below 200 $\mu$W, which is enough to align a triplet of adjacent resonances. 
As the pump offset $\Delta_0 /2 \pi < -170$ GHz, the red $\omega_-$ and blue $\omega_+$ signals emerge from noise (-73 dBm = 50 pW), 
as can shown in Fig.\ref{fig:OPO}(a). When approaching -175 GHz, they abruptly increase by four orders of magnitude, a clear indication of an oscillation threshold. Interestingly, the pump offset, is, to a good approximation directly proportional to the energy stored in the pump mode as the spectral shift is induced by linear absorption. Therefore, the result can be cast in a more familiar representation by estimating the equivalent pump power circulating in the cavity $P_{c,0}$ (see Supplementary Information section III.A) in Fig.\ref{fig:OPO}(b). Above threshold $P_{c,0}>P_{th}=175\mu W$ , the on-chip generated power increases linearly with $P_{c,0}-P_{th}$. More than $50\%$ of the excess pump power is converted. Strikingly, the threshold expected from our model is 170 $\mu$W (see Supplementary Table III). As the detuning is further increased, the parametric oscillation shuts off at $\Delta_0/2\pi$ = -183GHz, although the pump mode is still on resonance. The cavity is now misaligned. 

%
%
The  performance  of  our  ultracompact   system  based  on   interacting   standing   waves  is already comparable to that of recently demonstrated semiconductor microring and racetrack OPOs which exhibits power thresholds between 3 and 25 mW, as shown in Fig.\ref{fig:OPO}(c). The very low power threshold of our PhC OPO results from the strong confinement of the interacting modes, the large nonlinearity of semiconductors and a moderately large Q factor. The mode volumes of about $0.2\mu m^3$, are 150 times smaller than in ring resonators with comparable FSR \footnote{$V_m=30\mu m^3$ for FSR = 500 GHz\cite{pu2016} Considering the interaction volume $V_\chi=5.7\mu m^3$, this is still an order of magnitude smaller than $V_{\chi,ring}=1.54\times 2 \pi A_{eff} L=50\mu m^3$.}. So far, lower power thresholds were observed very recently\cite{chang2019} in AlGaAs based microring (36 $\mu$W), thanks to larger the nonlinearity of the material, much higher Qs and  larger FSR.  Lower values (down to 5 $\mu$W) are only reported in non-integrated and non-semiconductor resonators that require highly optimized fabrication process\cite{furst2010} and second order nonlinearity. Considering that current state-of-the art PhC cavities\cite{asano2017} exhibit $Q > 10^7$, OPOs with power thresholds below the $\mu$W level can be realistically considered. \\
As in any other laser source, power threshold is not the sole figure of merit. Also the overall energy  efficiency transfer from the pump to the side modes is crucial. In our case, the estimated total generated power, when considering outcoupling loss (7dB) is about 5 $\mu$W (measurement is in Supplementary Information section II.B), leading to a conversion efficiency which is about 2.5$\%$ of the coupled pump power. This value can be compared to the one of 17 $\%$ very recently reported in AlN ring resonators\cite{Bruch2019}, which are operated at much larger power (10 mW) and exploit the $\chi^{(2)}$ nonlinearity. However, the PhC OPO measured slope efficiency (signal+idler) is above $50\%$, indicating that a much better overall efficiency is possible by optimizing the coupling to the feeding waveguide.\\
Finally, it is interesting to point out that our OPO behaves as a pure degenerate parametric system with only three cavity modes interacting. Simultaneous alignment of more than one triplet is very unlikely. This is apparent in Fig. \ref{fig:cavity_tuning}(e), where another triplet ($x$, $+$ and $0$) is aligned at a different pump offset $\Delta_0\approx 50$ GHz. This characteristic, possible in ring resonator with a sophisticated sidewall corrugation\cite{luSrinivasan2019}, could find an application in laser noise reduction\cite{matsko2019}.For the same reason, resonant nonlinear contributions such as Raman can be ignored, unless the resonator is deliberately designed for, as in Ref. \cite{takahashi2013}.\\
In conclusion, we achieved a low-power operating OPO with a footprint far smaller than other competing approaches.  It is of particular interest when it comes to considering incorporation within sophisticated optical circuits\cite{Marty2019}. This demonstration opens up exciting avenues for building an integrated all semiconductor platform to optically generate and process both classical and quantum data.  
\section*{Methods}

The cavities are suspended membranes made of Indium Gallium Phosphide lattice-matched to GaAs. The photonic crystal is created using e-beam lithography, dry etch of the hard mask, inductively coupled plasma etching of the holes and wet etching of the underlying material substrate to release the membrane\cite{combrie2009}. The large electronic gap (1.89 eV) prevents two-photon absorption when operating the telecom spectral band, while the residual absorption rate $\Gamma_{abs}=2\times10 GHz$ is very low\cite{ghorbel2019}. Optical measurements have been performed on a temperature stabilized sample holder position stage and the two-way optical access is provided by a microscope objective lensed fiber, actuated by a 3-axis nanopositionner stage, and a circulator. A tunable laser source is connected to the sample through a variable attenuator and the output is directly fed to the Optical Spectrum Analyzer.

\newpage



\newpage
\vspace*{5ex}
\noindent\textbf{Supplementary Information}  is available in the online version of the paper.\\
\newpage
%
%

%
%
\newpage

\newpage

\end{document}